\documentclass{ieeeaccess}
\usepackage{cite}
\usepackage{amsmath,amssymb,amsfonts}
\usepackage{algorithmic}
\usepackage{graphicx}
\usepackage{textcomp}
\def\BibTeX{{\rm B\kern-.05em{\sc i\kern-.025em b}\kern-.08em
    T\kern-.1667em\lower.7ex\hbox{E}\kern-.125emX}}
    
\usepackage{cite}
\usepackage{graphicx,bm,color,hyperref}   
\usepackage{diagbox}
\usepackage{subcaption}
\usepackage{xcolor}
\renewcommand{\baselinestretch}{1.05}

\newtheorem{theorem}{Theorem}

\newtheorem{lemma}[theorem]{Lemma}

\usepackage{tabularx}
\usepackage{booktabs}  
    
\usepackage{mathrsfs}
\usepackage{dsfont}
\usepackage{mathtools}
\usepackage{bbm,bm}
\usepackage{mathdots}
\usepackage{amssymb}
\usepackage{amsfonts}
\newtheorem{Proposition}{Proposition}
\newtheorem{Corollary}{Corollary}
\newcommand\AME{\mathrm{AME}}
\newcommand\dH{d_H}
\newcommand\dC{d}
\newcommand\dloc{q}
\DeclareMathOperator{\wt}{wt}
\DeclareMathOperator{\lspan}{span}
\DeclareMathOperator{\hiH}{\mathcal{H}}
\DeclareMathOperator{\C}{\mathscr{C}}
\DeclareMathOperator{\Cq}{\mathcal{C}}
\DeclareMathOperator{\e}{\mathrm{e}}
\DeclareMathOperator{\iu}{\mathrm{i}}
\DeclareMathOperator{\Tr}{\mathrm{Tr}}
\newcommand{\ket}[1]{\vert #1 \rangle}
\newcommand{\bra}[1]{\langle #1 \vert}
\newcommand{\ketbra}[2]{\vert #1 \rangle\langle #2\vert}
\newcommand{\braket}[2]{\langle #1 \vert #2 \rangle}
\newcommand{\ceil}[1]{\left\lceil #1 \right\rceil}
\newcommand{\floor}[1]{\left\lfloor #1 \right\rfloor}
\newcommand{\1}{\ensuremath{\mathbbm{1}}}
\newcommand{\kq}{\tilde{k}}
\newcommand\uni {\mbox{-}\mathrm{UNI}}
\newcommand\remove[1]{\widehat{#1}}
\newcommand\modM{\mathcal{M}} 
\newcommand{\Sh}[1]{\overline{#1}}

\renewcommand{\emph}{\textit}
\newcommand\dq{d} 
\newcommand\lX{\overline{X}} 
\newcommand\lZ{\overline{Z}} 

\begin{document}
\history{Date of publication xxxx 00, 0000, date of current version xxxx 00, 0000.}
\doi{10.1109/ACCESS.2017.DOI}

\title{Modifying method of constructing quantum codes from highly entangled states}

\author{\uppercase{Zahra Raissi}}

\address{ICFO-Institut de Ciencies Fotoniques, The Barcelona Institute of Science and Technology, Castelldefels (Barcelona), 08860, Spain (e-mail: z.raissi2@gmail.com)}

\begin{abstract}
There is a connection between classical codes, highly entangled pure states (called $k$-uniform or absolutely maximally entangled (AME) states), and quantum error correcting codes (QECCs).
This leads to a systematic method to construct stabilizer QECCs by starting from a $k$-uniform state or the corresponding classical code and tracing out one party at each step.
We provide explicit constructions for codewords, encoding procedure and stabilizer formalism of the QECCs by describing the changes that partial traces cause on  the corresponding generator matrix of the classical codes. 
We then modify the method to produce another set of stabilizer QECCs that encode a logical qudit into a subspace spanned by AME states. 
This construction produces quantum codes
starting from an AME state without tracing out any party.
Therefore, quantum stabilizer codes with larger codespace can be constructed.
\end{abstract}


\titlepgskip=-15pt

\maketitle

\section{Introduction}
\label{sec:introduction}
\PARstart{Q}{uantum} error correction is one of the main challenges in the field of quantum computation and one of our attempts to use multipartite entangled states in applications \cite{Steane96-multipartite-QECC,Rains}.
Investigation of the connection between stabilizer quantum codes and existing classical error correcting codes led us to understand the structure of quantum codes and their connection to the highly entangled subspaces \cite{Steane96,Gottesman,Calderbank1998,Scott2004,ZCAA}.
In particular, there are two well-known methods for constructing stabilizer QECCs.
First, a general framework is constructing QECCs from known classical codes and the associated entangled states \cite{Gottesman-thesis,ZCAA}. 
The second method is constructing new codes from existing ones \cite{Gottesman-thesis,Calderbank1998}.

It is now well understood that a particular type of highly entangled pure states, called $k$-uniform states, represent stabilizer QECCs \cite{Scott2004}.
$k$-uniform (or for simplicity $k \uni$) states are genuinely entangled, which refers to the fact that all subsystems of size $k$ are correlated and that the states are not separable concerning any possible splitting of $k$ subsystems. 
Those that are maximally entangled along with any splitting the parties into two groups are called AME i.e., $\floor{n/2} \uni$ states.
As shown in \cite{ZCAA,ZACA,Helwig-graphstates}, one method of constructing these states is based on the connection between them and a family of classical error correcting codes known as maximum distance separable (MDS) codes.
This represents a method of constructing QECCs from classical codes (CSS construction, see \cite{Gottesman}). 
Moreover, the extra knowledge on code parameters of classical codes provides a great advantage to construct
quantum codes \cite{Gottesman-thesis}.

The second method that simplifies the task of finding quantum codes is to use existing codes to construct new ones.
Implementing some modification techniques on a given code can produce new code with different parameters \cite{Gottesman-thesis,Calderbank1998}. 
A non trivial manipulation is to remove
the last party of a given stabilizer code and convert it into a new code with one fewer party.
In this method, the derived code from a given stabilizer code is a stabilizer code \cite{Rains}.

Combining the two methods leads to a family of QECCs.
With this technique, one starts from a $k \uni$ state of $n$ parties and constructs the QECC by taking partial trace over one particle, i.e., generates a code with $n-1$ parties.
Repeating this technique produces a family of QECCs with a different set of code parameters \cite{FelixMarkus,Daniel}.
This method can be called \emph{Shortening} which refers to the connection it has with the classical codes and the subspace of constructed quantum codes which are spanned by the highly entangled states \cite{Calderbank1998,Ketkar,Markus-Roetteler,Sarvepalli}. 

So far, in the previous literatures, the description of the stabilizer formalism of the codes constructed from the Shortening process was in the center of attention \cite{Gottesman-thesis,Daniel}.
There, in order to construct the quantum codes one needs to find the generators of stabilizer of the state
in the way that one of them ends to $X$, one other ends to $Z$ and the rest of the stabilizers end $\1$, for details see\cite{Gottesman-thesis}.
This could be done more easier for the binary codes but for higher local dimension $\dloc$ finding this specific pattern needs effort.
Also, we aim at using quantum codes in quantum communication and computation applications, therefore, we need to know the form of the codewords and logical operators.

In this paper, we work on constructing a set of QECCs starting from $k \uni$ states. 
We start from a $k \uni$ state constructed from classical codes and describe how the partial trace over a particle change the corresponding generator matrix of the classical code.
Using this, unlike previous construction, we present the list of the codewords and logical operators as well as presenting the stabilizer formalism.
We also discuss the structure of the highly entangled subspace of the quantum stabilizer codes.

Moreover, we introduce a new systematic way of constructing quantum codes from existing ones. 
This method that we call \emph{modified-Shortening}, produce QECCs without removing any party (without taking the partial trace). 
Therefore, quantum stabilizer codes with larger codespace can be constructed that improve the Shortening procedure.
For this, we start from an AME state and without removing any party we construct a QECC whose codewords are all AME states. 
More precisely, starting from an AME
state or alternatively the quantum code $[\![n, 0,\floor{n/2}+ 1]\!]_\dloc$, we show how to produce a family
of quantum error correcting codes $[\![n,1,\floor{n/2}]\!]_\dloc$, while in the previous method one could only construct stabilizer QECC with parameters $[\![n-1, 1, \floor{n/2}]\!]_\dloc$.
We present the codewords and stabilizer formalism.

The structure is as follows: 
in Section \ref{sec:Sh-process} we discuss the connection between classical codes, $k \uni$ states, and QECCs.
Then, we discuss the Shortening process, the method of constructing a family of QECCs by taking partial trace over parties of those $k \uni$ states.
In Section \ref{sec:Explicit-Shortening} we describe the explicit constructions of QECCs using the Shortening process.
We describe how the partial trace changes the generator matrices of the corresponding classical codes used to construct the initial $k \uni$ states. 
Considering this, we then present the closed form expression of codewords and logical operators of the QECCs as well as presenting the stabilizer formalism.
In Section \ref{sec:mod-sh} we show how the modified-Shortening can produce QECCs starting from an AME state without taking any partial trace.

\section{Classical codes, \lowercase {$k$}-uniform states and optimal quantum codes}
\label{sec:Sh-process}
We begin with a short review of general definitions of classical linear error correcting codes, $k \uni$ states and quantum error correcting codes.
Then we describe how these concepts lead us to construct a family of QECCs from those $k \uni$ states that are constructed from classical linear codes.

{\it Classical error correcting codes.}
A linear classical code $\C=[n,k,\dH]_\dloc$ consists of $\dloc ^k$ codewords with length $n$ over $\dloc$-level dits.
The \emph{Hamming distance} $\dH$ between two codewords is defined as the number of positions in which they differ, and it guarantees to recover the original message if noise causes $t=\floor{(\dH -1)/2}$ dits of the code. 
The \emph{Singleton bound} provides upper bound on the maximally achievable minimal Hamming distance between any two codewords 
\begin{equation}\label{eq:singleton}
 \dH \leq n-k+1\ .
\end{equation}
In general, for the encoding procedure of the linear code $\C =[n,k,\dH]_\dloc$, it is possible to define a \emph{generator matrix} $G_{k \times n}$, in which the codewords are all possible linear combinations of the rows of such matrix.
And, it can always be written in the standard form \cite[Chapter~1]{MacWilliams}
\begin{equation}\label{eq:generator-standard}
 G_{k\times n}=[\1 _k | A] \ ,
\end{equation}
where $\1_k$ is identity matrix with size $k \times k$, and $A \in GF(\dloc)^{k \times (n-k)}$.
Codes that satisfy the Singleton bound \eqref{eq:singleton}, refer to as maximum-distance separable (MDS) codes.
For a given MDS code any subset of up to $k$ columns of $G_{k\times n}$ are linearly independent \cite[Chapter~11]{MacWilliams}\cite{Singleton,Roth,Seroussi}.

{\it $k \uni $ states.} 
Pure states of $n$  distinguishable qudits, with this property that all $k$-qudit reductions of the whole system are maximally mixed are called $k \uni$ states.
A $k \uni$ state $\ket{\psi}$ in Hilbert space $\hiH(n,\dloc)=\mathbb{C}_\dloc^{\otimes n}$,  denote by $k \uni(n,\dloc)$, whenever the following conditions hold
\begin{equation}
\Tr_{S^c} \ketbra\psi\psi \propto \1 , \qquad
 \forall S \subset \{1,\ldots,n\}, \ |S| \leq k \,  ,
\end{equation}
where $S^c$ denotes the complementary set of $S$.
The minimal number of terms for which the condition of maximally mixed marginals can be fulfilled is $\dloc^k$, and the corresponding states called  \emph{$k \uni$ states of minimal support}.
Moreover, the Schmidt decomposition shows that a state can be at most $\floor{n/2} \uni$, i.e., $k\leq \floor{n/2}$.
The $\floor{n/2} \uni$ states are called absolutely maximally entangled state, or AME states for short.

A subclass of pure $k \uni$ states can be constructed by taking equally weighted superposition of all the $\dloc^k$ codewords of MDS codes $\C=[n,k,n-k+1]_\dloc$ with $k \leq \ceil{n/2}$ in the computational basis i.e.,
 \begin{equation}\label{eq:kuni-MDScode}
  \ket{\psi} = \sum_{\vec{v}\in {GF(\dloc)}^k} \, \ket{\vec{v}\, G_{k\times n}}\ .
 \end{equation} 
The stabilizer formalism is provided in Appendix \ref{app:stab-kuni}.
As a side remark, note that it is always possible to find a suitable generator matrix $G_{k \times n}$ to construct a $k \uni$ state over finite field $GF(\dloc)$, all one has to do is to take a power of a prime $\dloc$ sufficiently large \cite[Chapter~11]{MacWilliams}\cite{Roth,ZCAA,ZACA} (for the existence interval see \cite{Roth,ZACA}).

{\it Quantum error correcting codes.}
A \emph{pure stabilizer} QECC denotes by $\Cq=[\![n,\kq,d]\!]_\dloc$.
This code is a $\dloc ^{\kq}$ dimensional subspace of the Hilbert space $\hiH(n,\dloc)$, where $d$ is the minimum distance.
A QECC with minimum distance $d$ can correct errors that affect no more than $t=\floor{(d-1)/2}$ of the subsystems.
Moreover, similar to the classical Singleton bound, for QECC the \emph{quantum Singleton bound} for the code $\Cq$ relates the parameters $n$, $\kq$, and $d$ as follows, \cite{Knill-Laflamme,Rains}
\begin{equation}\label{eq:quantum-singleton-bound}
  d \leq \frac{n-\kq}{2}+1 \ .
\end{equation}
A quantum code with maximum possible integer for $d$, is considered as an \emph{optimum code} (for further details see \cite[Definition~6]{Grassl-on-optimal}). 
We recall a strict definition of distance $d$, and the conditions under which a quantum code is a valid code in the next section.\\ 
 
The stabilizer codes are useful in applications of quantum computing and have proved to be particularly fruitful in understanding the structure of QECCs. 
There is a close connection between classical codes and stabilizer codes such that the construction of them can be reduced to that of the classical linear error correcting codes.
The knowledge on the code parameters of the classical codes provides a great advantage, and hence one simple method to construct the stabilizer codes is using the classical codes.
We know that $k \uni$ states of minimal support can be constructed from MDS codes, Eq.~\eqref{eq:kuni-MDScode}, and are a set of stabilizer QECCs \cite{Calderbank1998,Scott2004,Ketkar,Gottesman-thesis,Gottesman,ZCAA}.
In particular, a $k \uni$ state of $n$ particles corresponds to a pure QECC of distance $d=k+1$ and $\kq =0$, denotes by $\Cq=[\![n,0,k+1]\!]_{\dloc}$ \cite{Scott2004}.

Stabilizer quantum codes have also the property that they can be shortened, which means that the existence of a quantum pure stabilizer code (in this case a $k \uni$ state), implies the existence of a stabilizer QECC with larger code dimension whose spanning vectors are $(k-1) \uni$ states, see \cite[Theorem~6]{Calderbank1998}\cite[Lemma~70]{Ketkar} and \cite{Markus-Roetteler,Sarvepalli}.
Shortening process of the quantum stabilizer code $\Cq=[\![n, 0 ,k+1]\!]_\dloc$, yields the following code. \\

 \begin{Proposition}[Shortening]
    The existence of a pure stabilizer code $[\![n,\kq =0 ,d=k+1]\!]_\dloc$ associated to a $k \uni (n,\dloc)$ state guarantees the existence of a pure stabilizer QECC, $[\![n-1, 1,d-1=k]\!]_\dloc$ with a non-trivial subspace spanned by a set of $(k-1) \uni$ states.
 \end{Proposition}

\vspace*{0.3cm}

The Shortening process guarantees the construction of a new quantum code from an existing quantum code $\Cq=[\![n, 0 ,k+1]\!]_\dloc$ (or $k \uni$ state) with larger code dimension.
Repeating the Shortening process yields codes with $\kq >0$ and hence $\dloc ^{\kq} > 1$ dimensional subspace.

\section{Explicit construction of the Shortening process}\label{sec:Explicit-Shortening}  
In this section, after recalling the theory of the Shortening process, we discuss the conditions under which a subspace is a QECC, then we show explicitly how to find the codewords of the quantum codes using the Shortening procedure.

For this, unlike the previous works where one needs to find specific patterns for the stabilizers to compute the partial trace \cite{Gottesman-thesis}, we start from the generator matrix of a given $k \uni$ state and describe what the partial trace does on the corresponding generator matrix at each step.
The extra information on the code parameters of the classical error correcting codes used in each step of the Shortening process leads us to find explicit closed form expressions of the codewords and logical Pauli operators of the code.
By working out the details of the connection between classical codes and the codewords of the stabilizer quantum code, we find simpler implementations of constructing codespace.

More specifically, we construct the codespace of a quantum code $[\![n-r, r, d-r]\!]_\dloc$, with $r >0$, from a given $k \uni (n,\dloc)$ state, such that the subspace are spanned by $(k-r) \uni$ states of $n-r$ parties.
This construction requires $n -r \leq \dloc +1$ \cite{Grassl-on-optimal}, which refers to the existence condition of the classical MDS codes.
We find an explicit closed form expression of the codewords and the \emph{logical Pauli $\lX$ and $\lZ$ operators} of the code.

To define stabilizer quantum codes we first introduce some notations. 
Let's start with the definition of Pauli operators which act on a given local element of a product state $\ket j$ as follows
\begin{align}
 X\ket{j} &= \ket{j+1 \mod \dloc} \label{eq:defX}\\
 Z\ket{j} &= \omega^j\,\ket{j}\, , \label{eq:defZ}
\end{align}
with $\omega \coloneqq \e^{\iu 2 \pi/\dloc}$ the $\dloc$-th root of unity.
$X$ and $Z$ are unitary, traceless operators, and $X^\dloc = Z^\dloc = \1$.
For $a,b \in \{0,\dots, \dloc -1\}$ it holds that $\Tr(Z^a\,X^b) = \delta_{a,0}\,\delta_{b,0}$ and $Z\,X = \omega\,X\,Z$. 
We call operators that are tensor products of powers of Pauli operators \emph{Pauli strings}.

Let $\{\ket{\psi_m}\}_{m\in [\dloc ^{\kq}]}$ be a set of orthonormal quantum states of $n$ qudits spanning
a subspace $\Cq$.
We denote by $[\dloc ^{\kq}]$ a string of $\kq$ symbols that range from $0$ to $\dloc -1$, e.g., for the case that $\kq=1$ we have, $[\dloc] \coloneqq (0,\dots ,\dloc -1)$.
A stabilizer group $S$, contains an abelian subgroup of Pauli strings (exclude $-\1$), such that the non-trivial subspace $\Cq$ of $\hiH(n,\dloc)$ is stabilized by $S$.
A stabilized subspace $\Cq$ defines a quantum codespace as follows: 
\begin{equation}
 \Cq = \{\ket{\psi_m}\in \hiH(n,\dloc)\, \colon \ S_i\, \ket{\psi_m} = \ket{\psi_m} ,\, \forall S_i \in S\}.
\end{equation}
$S$ is generated by $n-\kq$ independent stabilizer operators $S_i$, so that the codespace $\Cq$ encodes $\kq$ logical qudits into $n$ physical qudits.
The code $\Cq$ with parameters $[\![n, \kq, d ]\!]_\dloc$ is a valid QECC if it obeys Knill-Laflamme conditions \cite{Knill-Laflamme,Ketkar}
\begin{equation} \label{eq:Knill-Laflamme}
  \forall m,m' \in [\dloc ^{\kq}] \colon\ \bra{\psi_m}E^\dagger F\ket{\psi_{m'}}=f(E^{\dagger}F)\,\delta_{m,m'} \ ,
\end{equation}
for all $E,F$ with $\wt(E^{\dagger}F) < \dq$.
Here, $\wt$ is the \emph{weight} of an operator which denotes the number of sites on which it acts non-trivially.
The parameter $d$ is the distance of the code,
which is the minimal number of local operations that act on single sites to create a non-zero
overlap between any two different states $\ket{\psi_m}$ and $\ket{\psi_{m'}}$, i.e.,
\begin{equation}
  \dC \coloneqq \min_{\ket{\psi},\ket{\psi'}\in \mathcal{C}, W} \{\wt(W) \colon 
  \bra{\psi}W\ket{\psi'}\ne 0 \land  \braket{\psi}{\psi'}= 0\}\, .
\end{equation}
Such a code can correct all errors that act non-trivially on up to $t\coloneqq\floor{(\dC-1)/2}$ physical qudits.

In the following we show how codes with larger code dimension $\kq >0$, can be obtained by the Shortening process.

\subsection{First step of Shortening and logical operators $\lX$ and $\lZ$:}
As the first step, the Shortening procedure can convert the code $[\![n, 0, k+1]\!]_\dloc$, into a code with parameters $[\![n-1, 1, k]\!]_\dloc$ such that a logical qudit of dimension $q$ is encoded in a $\dloc$-dimensional subspace spanned by $(k-1) \uni$ states.
In the following we show how to find the codespace $\Cq = \lspan(\{\ket{\psi_m}\}_{m\in [\dloc]})$. 
We start with the generator matrix $G_{k\times n}=[\1_k|A_{k\times (n-k)}]$ and remove the last row.
Because the generator matrix has the standard form, the $k$-th column of the resulting matrix contains only $0$s, so we remove this column too.
With these changes the generator matrix now transforms into a matrix of size $(k-1)\times (n-1)$
\begin{equation}\label{eq:one-Shortening}
 G_{k\times n}=[\1_k|A_{k\times (n-k)}] \
 \longrightarrow \ 
 G_{\remove{k}}=[\1_{k-1}|A_{(k-1)\times (n-k)}] \ ,
\end{equation}
where we denote by $G_{\remove{i}}$ the result of removing the $i$-th row and column from the original matrix $G$.
$G_{\remove k}$ contains $k-1$ linearly independent columns, 
therefore $G_{\remove k}$ is a valid generator matrix to construct an MDS code $\C=[n-1,k-1,n-k+1]_\dloc$  (while obviously every square submatrix of $A_{(k-1)\times (n-k)}$ is non-singular).
Hence, the state 
 \begin{equation}\label{eq:psi0}
   \ket{\psi _0} = \sum_{\vec{v}\in {GF(\dloc)}^{k-1}} \, 
   \ket{\vec{v}\ G_{\remove{k}}}\ ,
 \end{equation}
is a $(k-1) \uni$ state.
Now, we define the operator $M$ which is a string of powers of $X$ where the exponents correspond to the removed last row of matrix $G$, concretely
 \begin{equation}\label{eq:M-Shortening} 
   M \coloneqq 
  \underbrace{\1\otimes \dots \otimes \1}_{k-1} \otimes \underbrace{ X^{a_{k,1}} \otimes  X^{a_{k,2}} \otimes  \dots \otimes X^{a_{k,(n-k)}}  }_{n-k} \ .
 \end{equation}
where we denoted elements of matrix $A_{k\times (n-k)}$ by $a_{i,j}$.
In the following lemma, we show how the Pauli string $M$, defines a QECC made of $(k-1) \uni$ states. \\

 \begin{lemma} \label{lemma:Sh.}
   Consider the $(k-1) \uni$ state $\ket{\psi_0}$, Eq.~\eqref{eq:psi0}, constructed from the generator matrix $G_{\remove k}$ and the $M$ operator constructed from the elements of the last row of the $G_{k\times n}$ matrix, Eq.~\eqref{eq:M-Shortening}.
  The subspace $ \Cq = \lspan(\{\ket{\psi_m}_{m\in [\dloc]}\}) \subset \mathbb{C}_{\dloc}^{\otimes {n-1}} $ with 
  \begin{equation}\label{eq:psi_m-Shortening}
   \ket{\psi_m} = M^m \ket{\psi_0}  \qquad 0 \leq m \leq \dloc-1  ,
  \end{equation} 
 is a QECC with parameters $[\![n-1,1,k]\!]_\dloc$.
 \end{lemma}
 
\vspace*{0.3cm}

Before providing the proof let us present an example.
As an example we go through the construction of optimal quantum code $[\![3,1,2]\!]_3$ from $\AME(4,3)$ state (or alternatively quantum code $[\![4,0,3]\!]_3$).
The generator matrix $G_{2 \times 4}$ that constructs $\AME(4,3)=\sum_{i,j \in GF(3)} \ket{i,j,i+j,i+2j}$, and the reduced generator matrix $ G_{\remove{2}}$ are given by
 \begin{equation}
  G_{2\times 4} 
   =\left[  \begin{array}{cc|cc}
1 & 0 & 1 & 1 \\
0 & 1 & 1 & 2
 \end{array} \right]
 \ \longrightarrow \ 
 G_{\remove{2}} 
   =\left[  \begin{array}{c|cc}
1 & 1 & 1 \\
 \end{array} \right]\ .
 \end{equation}
The reduced generator matrix $G_{\remove{2}}$ yields the following closed form expression for one of the codewords
 \begin{equation}
   \ket{\psi_0} = \sum_{\vec{v} \in GF(3)} \ket{\vec{v} \, G_{\remove{2}}} = \sum_{i \in GF(3)} \ket{i,i,i} \ ,
 \end{equation}
while the other codewords can be obtained by performing the powers of $M$ operator, that have the form
 \begin{equation}\label{eq:M-[[3,1,2]]}
  M  = \1 \otimes X \otimes X^2 \ ,
 \end{equation}
on the state $\ket {\psi_0}$.
Using this formalism, the states
 \begin{equation}\label{eq:codewords[[3,2,1]]}
   \begin{split}
    \ket{\psi_0} &= \sum_{i} \ket{i, i, i} \\
    \ket{\psi_1} &= M \ket{\psi_0} = \sum_{i} \ket{i, i+1, i+2} \\
    \ket{\psi_2} &= M^2 \ket{\psi_0} = \sum_{i} \ket{i, i+2, i+1} \ ,
   \end{split}
 \end{equation}
list all of the codewords of the QECC $[\![3,1,2]\!]_3$ that is constructed from $\AME(4,3)$.
One can also check that the superposition of these codewords with arbitrary coefficients $0 \leq \alpha_i \leq 1$ 
 \begin{equation}
  \ket{\Psi} = \alpha_0 \ket{\psi_0} + \alpha_1 \ket{\psi_1} + \alpha_2 \ket{\psi_2} \ ,
 \end{equation}
is a $1 \uni$ state the same as for each codeword (to study more about entangled subspaces see \cite{completely-entangled-subspaces}).
 \begin{IEEEproof} [Proof of Lemma~\ref{lemma:Sh.}]
   First of all we need to show that any two codewords from the code state space are orthogonal. 
  To do this, we recall that all rows of the $G_{k \times n}$ matrix and every linear combination of them form codewords with specific Hamming distance, $\dH=n-k+1$ \cite[Chapter~11]{MacWilliams}.
  The state $\ket{\psi_0}$ and operator $M$ are formed by combination of the rows of the $G$ matrix after removing the $k$-th column.
  Performing the operator $M^m$ for a given $m \in [\dloc]$ on the state $\ket{\psi_0}$ is the same as adding a specific codeword (linear combination of the last row of $G_{k \times n}$) to the set of codewords that form the state $\ket{\psi_0}$.
 Hence, performing two different $M^m$ and $M^{m'}$ operators for all $m,m'\in [\dloc]$ on the state $\ket {\psi_0}$ produce two states such that
   \begin{align}\label{eq:orthogonal-codewords}
    & \braket{\psi_m}{\psi_{m'}} = \delta_{m,m'}\\
    & \bra{\psi_m} W \ket{\psi_{m'}} = 0 \label{eq:distance-codewords}\ ,
   \end{align} 
  where $\wt (W) < \dH-1 = n-k$.
  Note that all the states $\ket{\psi_m}$ are $(k-1) \uni$, as acting with local unitaries does not change the entanglement properties.
  
   The codespace $\Cq=\lspan\{\ket{\psi_m}\}$ is a QECC if and only if it satisfies two conditions.
   (i) In the presence of errors one should be able to distinguish two different codewords  \cite{Gottesman,Gottesman-thesis}. 
  Considering this, Eq.~\eqref{eq:distance-codewords} implies that the minimum number of single-qudit operations that are needed to create a non-zero overlap between any two orthogonal states is $n-k$.
  Therefore, for all errors $E$ and $F$ with weight such that $1 <\wt(E^{\dagger}F)<n-k$ and all $m,m'\in [\dloc]$ with $m\neq m'$ we have 
   \begin{equation}
    \bra{\psi_m} E^{\dagger} F\ket{\psi_{m'}} =0 \ .
   \end{equation}
 The above condition is a direct consequence of the fact that the minimum distance between two different codewords $\ket{\psi_m}$ and $\ket{\psi_{m'}}$ is $\dH-1= n-k$.
  (ii) 
  In addition, one should also be able to distinguish different errors when they act non-trivially on a given codeword $\ket{\psi_m}$ (see Eq.~(28) of  \cite{Gottesman}).
  As the states $\ket{\psi_m}$ for every $m\in [\dloc]$, are $(k-1) \uni$ state, then one gets
  \begin{equation}
   \bra{\psi_m} E^{\dagger} F \ket{\psi_m} = \Tr( E^{\dagger} F ) = 0 \ , \qquad \forall m\in [\dloc]\ .
  \end{equation}
  for errors $ E^{\dagger} F$ that act non-trivially on any subset of less than $k-1$ sites, i.e., $\wt(E^\dagger F) < k-1$.
  As we always have $n-k \geq k-1$, then, considering the conditions (i) and (ii) the code distance is $d=\min(n-k+1,k)=k$.
  By the definition Eq.~\eqref{eq:Knill-Laflamme}, one can conclude that the subspace $\Cq$ is a $[\![n,1,d=k]\!]_\dloc$ QECC.
\end{IEEEproof}

As a side remark, note that if the $M$ operator, \eqref{eq:M-Shortening}, contains only $k$ of the $X$ operators with the vector of exponents described above, the distance of the code is still $d=k$.
The stabilizer formalism is presented in Appendix \ref{app:stab-Shortening}. 

Now that the codewords and the code distance are determined, we can find logical $\lX$ and $\lZ$ operators. 
Logical Pauli operators are unitary operators which act non-trivially on the codeword space or logical states $\ket{\psi_m}$, but preserve it.
The logical Pauli operators $\lX$ and $\lZ$ preserve the codeword space since they commute with all the stabilizer generators.
The logical operators anti-commute with each other.
In the theory of quantum error correcting codes, it can always be useful to express $\lX$ and $\lZ$ in terms of their action as Pauli operators $X$ and $Z$ on the physical qudits used in codespace. 

In the Shortening construction, the logical qudits are the set of the states $\{\ket{\psi_m}\}_{m \in [\dloc^k]}$, Eq.~\eqref{eq:psi_m-Shortening}, which are constructed by performing the powers of $M$ operator on the state $\ket{\psi_0}$.
This shows that the $M$ operator is the logical $\lX$ Pauli operator that by starting from the logical qudit $\ket {\psi_0}$, Eq.~\eqref{eq:psi0}, it can construct entire codespace
 \begin{equation}
  \lX^m \ket{\psi_0} \equiv M^m \ket{\psi_0} = \ket{\psi_m} \ ,
 \end{equation}
for all $m \in [\dloc^m]$.
In this construction, we started from a $k \uni$ state constructed from an MDS code that has stabilizers involving $X$ or $Z$ operators, namely $X$ stabilizers or $Z$ stabilizers respectively (see Appendix~\ref{app:stab-kuni} for more details). 
The $X$ stabilizers are powers of $X$ where the exponents corresponding to every row of the generator matrix $G_{k\times n}=[\1_k | A]$, and the $Z$ stabilizers are $Z$ Pauli string that the exponents correspond to the rows of  $H=[A^T | \1_{n-k}]$ known as parity check matrix (for more details see appendix~\ref{app:stab-kuni} as well as \cite{ZCAA}).
After eliminating one row and column of the $G$ matrix to construct the $M$ (or logical $\lX$) operator, the stabilizer that corresponds to the last row of the $H$ matrix does not commute with $M$ any more while it commutes with the rest of the stabilizers.
Therefore, $\lZ$ is a string of powers of $Z$ where the exponents correspond to the last row of the parity check matrix $H_{(n-k) \times k}=[A^T | \1_{n-k}]$ when the $k$-th column is removed, i.e.,
 \begin{equation}\label{eq:lZ-Shortening} 
 \begin{split}
   \lZ & \coloneqq \\
  & \underbrace{ Z^{-a_{1,(n-k)}} \otimes Z^{-a_{2,(n-k)}} \otimes \dots \otimes Z^{-a_{(k-1),(n-k)}}}_{k-1} \otimes  \\
  & 
   \underbrace{ \1\otimes \dots \otimes \1 \otimes Z}_{n-k} \ .
  \end{split}
 \end{equation}
In the example we considered before the codewords of the quantum code $[\![3,1,2]\!]_3$ are presented in Eq.~\eqref{eq:codewords[[3,2,1]]}, and the $M$ operator \eqref{eq:M-[[3,1,2]]} is the $\lX$ Pauli operator.
In this construction
 \begin{equation}
   \begin{split}
    \lX &= M = \1 \otimes X \otimes X^2 \\
    \lZ &= Z^{-1} \otimes \1 \otimes Z \ ,
   \end{split}
 \end{equation}
are the logical Pauli operators of the QECC $[\![3,1,2]\!]_3$.

\subsection{Second step of Shortening and logical operators $\lX$ and $\lZ$:}
The second step of the Shortening procedure converts the code $[\![n-1, 1, k]\!]_\dloc$, into a code with parameters $[\![n-2, 2, k-1]\!]_\dloc$ with a $\dloc^2$-dimensional subspace spanned by $(k-2) \uni$ states.
For that we proceed in a similar way and remove the $k-1$ and the $k$-th columns and rows of the original generator matrix
\begin{equation}\label{eq:two-Shortening}
 \begin{split}
 G_{k\times n} &= [\1_k|A_{k\times (n-k)}] \\
 \longrightarrow \ 
 G_{\remove{k-1} , \,  \remove{k}} &= [\1_{k-2}|A_{(k-2)\times (n-k)}] \ .
  \end{split}
\end{equation}
The structure is the same as the first step, and hence, it is obvious that $ G_{\remove{k-1} , \,  \remove{k}}$ is the generator matrix of an MDS code $[n-2 , k-2, n-k+1]_\dloc$.
Therefore, $(k-2) \uni$ state $\ket{\psi _{00}}$ can be constructed via
 \begin{equation}\label{eq:psi00}
   \ket{\psi _{00}} = \sum_{\vec{v}\in {GF(\dloc)}^{k-2}} \, 
   \ket{\vec{v}\ G_{\remove{k-1} , \,  \remove{k}}}\ .
 \end{equation}
Two Pauli strings $M_1$ and $M_2$ that involve $X$ operators can be defined such that the vector of exponents are the $k-1$ and the $k$-th rows of matrix $G_{k \times n}$ while both the $k-1$ and $k$-th columns are removed:
   \begin{equation}\label{eq:M1-Shortening}
     \begin{split}
     & M_1  \coloneqq \\
  & \underbrace{\1\otimes \dots \otimes \1}_{k-2} \otimes \underbrace{ X^{a_{k-1,1}} \otimes  X^{a_{k-1,2}} \otimes  \dots \otimes X^{a_{k-1,(n-k)}}  }_{n-k} 
      \end{split}
   \end{equation}
   \begin{equation} \label{eq:M2-Shortening}
     \begin{split}
    & M_2 \coloneqq \\
  & \underbrace{\1\otimes \dots \otimes \1}_{k-2} \otimes \underbrace{ X^{a_{k,1}} \ \ \, \otimes  X^{a_{k,2}}  \ \ \, \otimes  \dots \otimes X^{a_{k,(n-k)}}  }_{n-k} \ .
    \end{split}
   \end{equation}
Finally, the codespace $\Cq$ is spanned by the $(k-2) \uni$ states 
 \begin{equation}
   \ket{\psi_{m_1,m_2}} = M_1^{m_1} M_2^{m_2} \ket{\psi_{00}}
   \qquad  0 \leq m_1, m_2 \leq \dloc-1 \ .
 \end{equation}
By the same argument as above, the fact that the state $\ket{\psi_{00}}$ and operators $M_1$ and $M_2$ are linear combination of the rows of the matrix $G_{k\times n}$ (or codewords of the MDS code $[n,k,n-k+1]_\dloc$), where two parties are removed, leads us to the first Knill-Laflamme condition (i):
 \begin{align}
  & \braket{\psi_{m_1,m_2}}{\psi_{m'_1,m'_2}} = \delta_{m_1,m'_1}\delta_{m_2,m'_2} \\
  & \bra{\psi_{m_1,m_2}} E^\dagger F \ket{\psi_{m'_1,m'_2}} = 0 \ ,
 \end{align}
with $\wt(E^\dagger F) < \dH-2=n-k-1$.
The second condition can be satisfied because:  (ii) the subspace $\Cq$ is manifestly spanned by orthogonal $(k-2) \uni$ states $\ket{\psi_{m_1,m_2}}$ so that, 
 \begin{equation}
  \bra{\psi_{m_1,m_2}} E^\dagger F \ket{\psi_{m_1,m_2}} = \Tr(E^\dagger F) =0 \ ,
 \end{equation}
 for $\wt(E^\dagger F) < k-2$.
This implies that the code distance of the QECC with codespace $\Cq=\lspan\{\ket{\psi_{m_1,m_2}}\}$ is $d=\min(n-k,k-1)=k-1$.

The two operators $M_1$ and $M_2$, Eqs.~\eqref{eq:M1-Shortening} and \eqref{eq:M2-Shortening}, are the logical $\lX$ operators for the two qudits of the code.
And the logical $\lZ$ operators are Pauli strings of powers $Z$ where the exponents correspond to rows $n-k$- and $n-k-1$-th of matrix $H$.
More specifically  the two operators
    \begin{equation}\label{eq:Z1-Shortening}
    \begin{split}
     & \lZ_1 \coloneqq \\
  & \underbrace{ Z^{-a_{1,(n-k-1)}} \otimes Z^{-a_{2,(n-k-1)}} \otimes \dots \otimes Z^{-a_{(k-1),(n-k-1)}}}_{k-1} \otimes 
  \\ &
  \underbrace{ \1\otimes \dots \otimes Z \otimes \1}_{n-k} 
  \end{split}
     \end{equation}
     
     \begin{equation}
         \begin{split}
    & \lZ_2 \coloneqq  \\
    & \underbrace{ Z^{-a_{1,(n-k) \phantom{-1} }} \otimes Z^{-a_{2,(n-k) \phantom{-1}}} \otimes \dots \otimes Z^{-a_{(k-1),(n-k) \phantom{-1}}}}_{k-1} \otimes
    \\ &    
     \underbrace{ \1\otimes \dots \otimes \1 \otimes Z}_{n-k}  \label{eq:Z2-Shortening}\ ,
    \end{split}
   \end{equation}
commute with all the stabilizers, yet anti-commute with $M_1$ and $M_2$.

As an example of our construction, we start with $\AME(6,5) = \sum_{\vec{v} \in GF(5)^3} \ket{\vec{v} \, G_{3 \times 6}}$ which is constructed from an MDS code $[6,3,4]_5$, over $GF(5)$ \cite{ZCAA}.
The generator matrix $G_{3\times 6}$ of the MDS code, and the reduced generator matrix $\Sh G_{2\times 5}$ are 
 \begin{equation}
 \begin{split}
  G_{3\times 6} 
   & =\left[  \begin{array}{ccc|ccc}
1 & 0 & 0 & 1 & 1 & 1 \\
0 & 1 & 0 & 1 & 2 & 3 \\
0 & 0 & 1 & 1 & 3 & 4  \\
 \end{array} \right] 
 \\
 \ \longrightarrow \ 
 G_{\remove{3}} 
   & =\left[  \begin{array}{cc|ccc}
1 & 0 & 1 & 1 & 1 \\
0 & 1 & 1 & 2 & 3 \\
 \end{array} \right]\ .
 \end{split}
 \end{equation}
Matrix $G_{\remove{3}} $ is the generator matrix of an MDS code $[5,2,4]_5$.
After the second step of reducing the generator matrix we get
 \begin{equation}
  \begin{split}
  G_{3\times 6} 
   & =\left[  \begin{array}{ccc|ccc}
1 & 0 & 0 & 1 & 1 & 1 \\
0 & 1 & 0 & 1 & 2 & 3 \\
0 & 0 & 1 & 1 & 3 & 4  \\
 \end{array} \right] \\
 \ \longrightarrow \ 
 G_{\remove{3},\remove{2}} 
   & =\left[  \begin{array}{c|ccc}
1  & 1 & 1 & 1 \\
 \end{array} \right]\ ,
  \end{split}
 \end{equation}
which is the generator matrix of the code $[4,1,4]_5$.
By taking the coherent superposition of its codeworks one has the $1 \uni$ state of $4$ parties
 \begin{equation}
  \ket{\psi_{00}} = \sum_{\vec{v}\in GF(5)} \ket{\vec{v} \, G_{\remove{3},\remove{2}} }
  =
   \sum_{i=0}^{4} \ket{i,i,i,i}\ .
 \end{equation}
 As it is discussed above in this case there are two operators
  \begin{align}
   M_1 &= \1 \otimes X \otimes X^2 \otimes X^3 \\
   M_2 &= \1 \otimes X \otimes X^3 \otimes X^4 \ .
  \end{align}
  States $ \ket{\psi_{m_1,m_2}} = M_1^{m_1} M_2^{m_2} \ket{\psi_{00}}$, where $0 \leq m_1, m_2 \leq \dloc-1$ form the subspace $\Cq =\lspan ({\ket{\psi_{m_1,m_2}}})$, which is a QECC with parameters $[\![4,2,2]\!]_5$.
And the logical $\lZ$ operators are 
  \begin{align}
   \lZ_1 &= Z^4 \otimes \1 \otimes Z \otimes \1 \\
   \lZ_2 &= Z^4 \otimes \1 \otimes \1 \otimes Z \ ,
  \end{align}
which are the last two rows of the parity check matrix $H_{3 \times 6}=G_{3\times 6}^T$ when the second and third columns are removed. 

In general, the Shortening procedure can be repeated $k-1$ times and codes with fewer particles and higher code dimension $\kq$ can be obtained (see Table \ref{table:Shortening-process}). 
Note that these codes can be optimal, saturating the quantum Singleton bound, only if $k=\floor{n/2}$. 
For that the starting point should be an AME state.

\begin{table}
\begin{center}
 \begin{tabular}{ccccccc} 
 Starting point: & & Step 1: \\
   $k \uni$ state $\equiv$ $[\![n,0,k+1]\!]_\dloc$ & $\longrightarrow$ &  $[\![n-1,1,k]\!]_\dloc$ &  $\longrightarrow$\\
  $1$-dimensional subspace & &  $\dloc$-dimensional subspace \\
    & &  $(k-1) \uni$ states 
\end{tabular}
\begin{tabular}{cccccccccc} 
$\hphantom{a}$ &  $\hphantom{a}$ & $\hphantom{a}$ &  $\hphantom{a}$ & $\hphantom{a}$ &  $\hphantom{a}$ & $\hphantom{a}$ & $\hphantom{a}$ &  \\
\end{tabular}
 \begin{tabular}{ccccccc} 
  Step 2:  & & &  & Step $r$: & & \\
  $[\![n-2,2,k-1]\!]_\dloc$ & $\longrightarrow$ & $\dots$ & $ \longrightarrow$ &  $[\![n-r,r,k-r+1]\!]_\dloc $  \\
  $\dloc^2$-dimensional subspace & & &  & $\dloc^r$-dimensional subspace & &  \\
  $(k-2) \uni$ states &  & &  & $(k-r) \uni$ states & & 
\end{tabular}
\begin{tabular}{cccccccccc} 
$\hphantom{a}$ &  $\hphantom{a}$ & $\hphantom{a}$ &  $\hphantom{a}$ & $\hphantom{a}$ &  $\hphantom{a}$ & $\hphantom{a}$ & $\hphantom{a}$ &  \\
\end{tabular}
\begin{tabular}{cccccccc} 
& & & Step $k-1$: \\
 $\longrightarrow$ & $\dots$ & $\, \longrightarrow$ & $[\![n-k+1,k-1,2]\!]_\dloc$    \\
& & & $\dloc^{k-1}$-dimensional subspace   \\
 & & & $1 \uni$ states   \\
\end{tabular}
\end{center}
 \caption{\label{table:Shortening-process} Shortening process: List of the stabilizer QECCs one can construct from a given $k \uni$ state.}
\end{table}

\section{Modified Shortening: Optimal quantum codes from AME states without tracing out particles} \label{sec:mod-sh}
The structure of the Shortening process is based on removing one particle (taking partial trace) from a given stabilizer QECC at each step.
In the previous section, we introduced a different view of the Shortening process that constructs stabilizer QECCs.
This view allows us to follow the knowledge on the code parameters of the corresponding classical code in each step of the Shortening, with this, we construct codewords and logical operators of the QECCs and study the structure of the codespace.

In this section, we improve the Shortening process to construct new codes from previous ones without tracing out any parties starting from an AME state.

We name this method modified-Shortening.
To introduce the method, we first review a systematic way of constructing generator matrices to construct classical MDS codes which provide explicit constructions and closed form expressions for AME states, i.e., $k=\floor{n/2} \uni$, or code $[\![n,0,\floor{n/2}+1]\!]_\dloc$.
Then we present the modified-Shortening which is a systematic method to construct QECC with parameters $[\![n,1,\floor{n/2} ]\!]_\dloc$ from an AME state defined by an MDS.

We start by introducing the concept of so-called \emph{Singleton arrays}, which is a special case of Cauchy matrix and have this property that all its square sub-matrices are non-singular \cite[Chapter~11]{MacWilliams}.
For any finite field $GF(\dloc)$, a Singleton array is defined to be
{\includegraphics[width=15.5in,height=1.82in,clip,keepaspectratio]{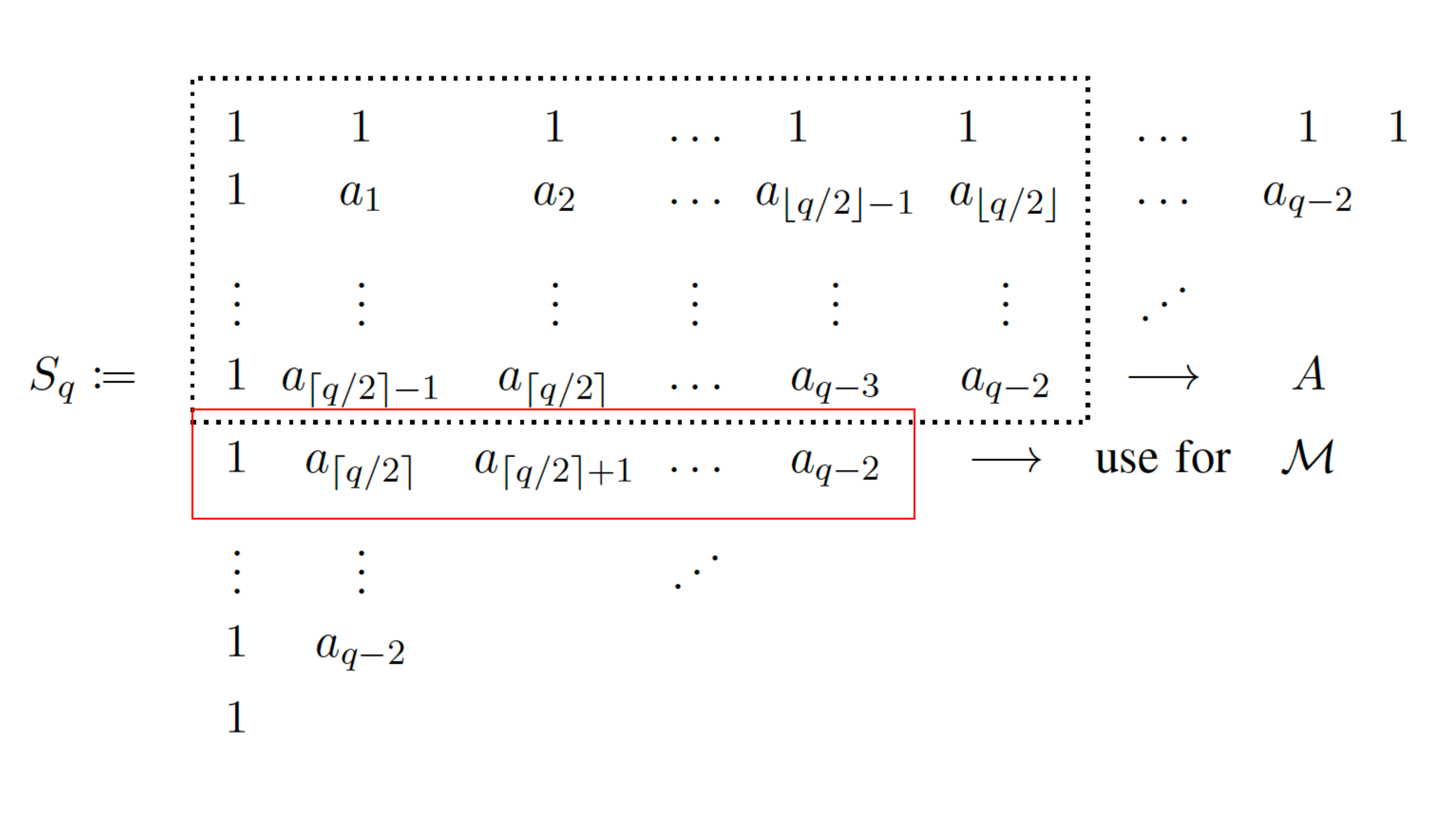}}
with
\begin{equation}
  a_i \coloneqq \frac{1}{1-\gamma^i} \ ,
\end{equation} 
where $\gamma$ is an element of $GF(\dloc)$ called primitive element.
It is shown that by taking the rectangular sub-matrix $A$, a suitable generator matrix $G =[\1 |A]$ for an MDS code can be constructed \cite[Chapter~11]{MacWilliams} \cite{ZCAA}.

The largest $A$ matrix this method constructs over $GF(\dloc)$ has size $\floor{\frac{\dloc+1}{2}} \times \ceil{\frac{\dloc+1}{2}}$.
By taking $n = \dloc+1$, one can construct the generator matrix $G_{\floor{n/2}\times n} =[\1|A]$ and  hence an MDS code with parameters $[n,\floor{n/2},\ceil{n/2}+1]_\dloc$.
The corresponding AME state $\ket{\phi_0}$ has the closed form expression \cite{ZCAA}
 \begin{equation}\label{eq:AME-phi0}
   \ket {\phi_0} = \sum_{\vec{v} \in GF(\dloc)^{\floor{n/2}}} \ket{\vec{v} \, G_{\floor{n/2} \times n}} \ .
 \end{equation}
Now, we consider the $\floor{(\dloc+1)/2} +1$-th row of the Singleton array $S_\dloc$ containing $\ceil{(\dloc+1)/2}-1$ elements $( 1,a_{\ceil{\dloc/2}}, a_{\ceil{\dloc/2}+1}, \dots , a_{\dloc-2} )$.
Using this, we define the Pauli string $\modM$ of length $n$, such that, the first $\floor{n/2}$ elements are identity matrices (as we have in each row of the generator matrix $G_{\floor{n/2} \times n}$), the vector of exponents of the $X$ operators is the $\floor{(\dloc+1)/2} +1$-th row of $S_\dloc$, and it contains one $Z$ operator as the $n$-th element, i.e., 
 \begin{equation}\label{eq:M-mod-Shortening} 
 \begin{split}
   & \modM \coloneqq  \\
   & \underbrace{\1\otimes \dots \otimes \1}_{\floor{\frac{n}{2}}} \otimes \underbrace{ X \otimes X^{a_{\ceil{\dloc/2}}} \otimes   X^{a_{\ceil{\dloc/2}+1}} \otimes \dots \otimes X^{a_{\dloc-2}}  }_{\ceil{\frac{n}{2}}-1} \otimes Z 
\ ,
 \end{split}
 \end{equation}
where $n=\dloc+1$.
In the following theorem we show that the AME states generated by acting with the Pauli string $\modM$ onto the state $\ket {\phi_0}$, Eq.~\eqref{eq:AME-phi0} from a QECC with parameters $[\![n,1,\floor{n/2} ]\!]_\dloc$.\\

 \begin{theorem}\label{theorem:modified-Shortening}
 From the AME state $\ket{\phi_0}$, or equivalently quantum code $[\![n,0,\floor{n/2}+1]\!]_\dloc$, Eq.~\eqref{eq:AME-phi0}, a QECC with parameters  $[\![n,1,\floor{n/2} ]\!]_\dloc$ can be constructed defined by the subspace $\Cq = \lspan (\{{\ket{\phi_m} _{m \in [\dloc]}} \}) \subset \mathbb{C}_{\dloc}^{\otimes n}$ with 
  \begin{equation}\label{eq:mod-Sh}
    \ket{\phi_m} \coloneqq \modM^m \ket{\phi_0} \qquad 0 \leq m \leq \dloc-1 \ .
  \end{equation}
 All $\ket{\phi_m}$ are AME states of $n$ parties with $n=\dloc+1$. 
 \end{theorem}

\vspace*{0.02cm} 
 
Before proceeding with the proof let us consider examples. 
In the previous chapter we discussed how to construct QECC $[\![3,1,2]\!]_3$ starting from $\AME(4,3)$ state  (or $[\![4,0,3]\!]_3$ code), and QECCs $[\![5,1,3]\!]_5$ and $[\![4,2,2]\!]_5$ from $\AME(6,5)$ state (or $[\![6,0,4]\!]_5$ code) using Shortening process. 
By using the modified-Shortening process one can construct  quantum codes $[\![4,1,2]\!]_3$ and $[\![6,1,3]\!]_5$ using $\AME(4,3)$ and $\AME(6,5)$ alternatively.
The codespace of code  $[\![4,1,2]\!]_3$ which are spanned by $\AME(4,3)$ states can be written as 
 \begin{equation}
  \begin{split}
   \ket{\phi_m} &= \modM^m \  \ket{\phi_0} \\
    &= \modM^m\sum_{i,j \in GF(3)} \ket{i,j,i+j,i+2j}
   \end{split}
       \qquad 0 \leq m \leq 2 \ ,
 \end{equation}
with 
 \begin{equation}
  \modM = \1 \otimes \1 \otimes X \otimes Z  \ .
 \end{equation}
And the codespace of QECC $[\![6,1,3]\!]_5$ that can be constructed with this method are presented by
 \begin{equation}
  \begin{split}
   & \ket{\phi_m} = \modM^m \  \ket{\phi_0} \\
    &= \modM^m\sum_{i,j,l \in GF(5)} \ket{i, j, l, i+j+l ,i+2j+3l, i+3j+4l} \ ,
   \end{split}
 \end{equation}
for $0 \leq m \leq 4$, while
 \begin{equation}
  \modM = \1 \otimes \1 \otimes \1 \otimes X \otimes X^4 \otimes Z \ .
 \end{equation}  

 \begin{IEEEproof}[Proof of Theorem~\ref{theorem:modified-Shortening}]
  The Codewords are produced by applying the $\modM^m$ operators for $m \in [\dloc]$ on the AME state $\ket {\phi_0}$.
 $\modM$ contains $X$ operators with the vector of exponents $\floor{(\dloc+1)/2} +1$-th row of $S_\dloc$ and one $Z$ operator.
 More explicitly, $\modM = M_X \, M_Z$ where,
 \begin{align} \label{eq:MX}
   M_X   \coloneqq &
  \underbrace{\1\otimes \dots \otimes \1}_{\floor{\frac{n}{2}}} \otimes \underbrace{ X \otimes X^{a_{\ceil{\dloc/2}}} \otimes   \dots \otimes X^{a_{\dloc-2}}  }_{\ceil{\frac{n}{2}}-1} \otimes \1
\\
   M_Z  \coloneqq &
   \underbrace{\1\otimes \1 \otimes \1 \dots \otimes \1 \otimes  \1 \otimes \1 \otimes   \dots  \otimes \1 \otimes \1 }_{n-1} \otimes Z \ . \label{eq:MZ}
 \end{align} 
 For the purpose of the proof we discuss how $M_X$ and $M_Z$ act on the state $\ket{\phi_0}$ separately.

  First, we show  that the application of $M_X^m$ for $m\in [\dloc]$ provides states with distance $\ceil{n/2}-1$.
  To do this, we take sub-matrix $A'_{ (\floor{\frac{\dloc+1}{2}}+1) \times (\ceil{\frac{\dloc+1}{2}}-1) }$
{\includegraphics[width=15in,height=1.87in,clip,keepaspectratio]{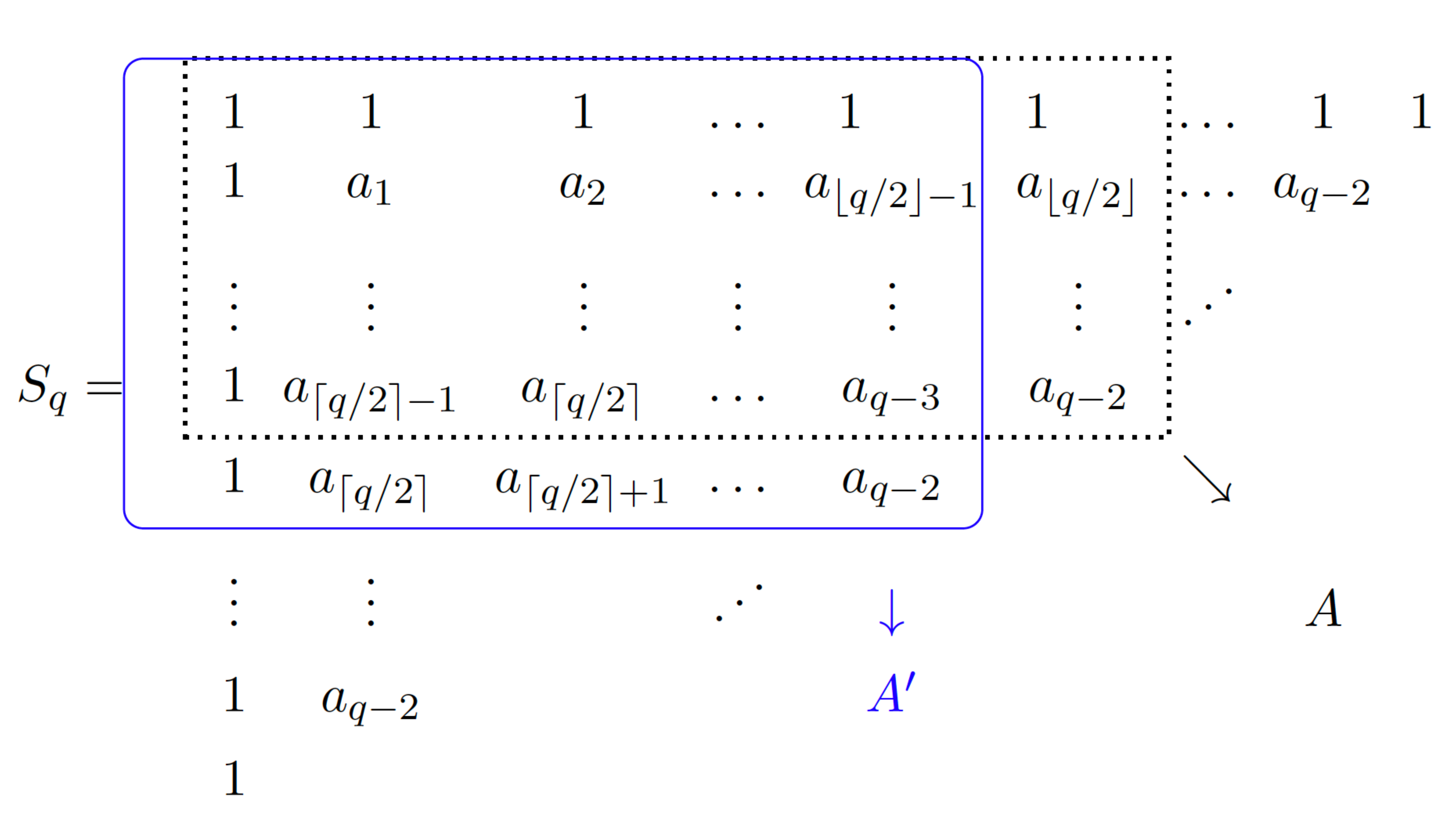}}
   of $S_\dloc$.
    Comparing it  with the largest submatrix $A$ of size ${\floor{\frac{\dloc+1}{2}} \times \ceil{\frac{\dloc+1}{2}}}$, the matrix $A'$ contains one more row and one less column.
Using $A'$, one can construct the generator matrix $G'_{(\floor{n/2}+1)\times n}=[\1_{\floor{n/2}+1}|A']$ and MDS code $\C'=[n,\floor{n/2}+1,\ceil{n/2}]_\dloc$.
  After one step of Shortening we get 
  \begin{equation}
  \begin{split}
     G'_{(\floor{n/2}+1)\times n} &= [\1_{\floor{n/2}+1}\, |\, A'_{ (\floor{\frac{\dloc+1}{2}}+1) \times (\ceil{\frac{\dloc+1}{2}}-1) }] \\
    \longrightarrow \ 
    G'_{\remove{\floor{\frac{n}{2}}+1}} &= [\1_{\floor{n/2}} \, | \, A'_{ \floor{\frac{\dloc+1}{2}} \times (\ceil{\frac{\dloc+1}{2}}-1) }]
 \ ,
 \end{split}
\end{equation}
  which is the generator matrix of an MDS code $[n-1,\floor{n/2},\ceil{n/2}]_\dloc$.
  Hence, an AME state $\ket{\psi'_0}$ of $n-1$ parties can be written as
  \begin{equation}
   \ket{\psi'_0} = \sum_{\vec{v} } \, \ket{ \vec{v} \, G'_{\remove{\floor{\frac{n}{2}}+1}}} \ .
  \end{equation}
  $G'_{\remove{\floor{\frac{n}{2}}+1}}$ is the same generator matrix as $G_{\floor{n/2} \times n} =[\1|A]$ if one deletes the last column, as well as the state $\ket{\psi'_0}$ is the same as state $\ket{\phi_0}$, Eq.~\eqref{eq:AME-phi0}, without considering the last party. 
As we discussed in the Shortening procedure, the $M$ operator that produces a subspace with specific distance $d$ is a string of only powers of the  $X$ operators with the vector of exponents defined by the last row of the generator matrix.
In this case, $M$ is the Pauli string of $X$ operators with the exponent vector $(\floor{(\dloc+1)/2} +1)$-th row of $S_\dloc$.
  Therefore, $M$ contains the first $n-1$ Pauli operators in $M_X$, Eq.~\eqref{eq:MX}.
  We can get the set of states
  \begin{equation}
    \ket{\psi'_{m}} = M^m \, \ket{\psi'_0} \qquad  0 \leq m \leq \dloc-1 \ .
  \end{equation}
    The same proof as that of Lemma~\ref{lemma:Sh.} establishes that the distance between every two different states of the above set of states is $\dH-1 = \ceil{n/2}-1$.
 This shows that because $\ket{\psi_m}$ is the same as $\ket{\phi_m}$ if one removes the last party, the distance between any two states $\ket{\phi_m}$ and $\ket{\phi_{m'}}$ is at least $d=\ceil{n/2}-1$ .

 Let us now consider the operator $M_Z$, Eq.~\eqref{eq:MZ}.
 For two different powers $m$ and $m'$, performing $M_Z^m$ and $M_Z^{m'}$  on the state $\ket{\phi_0}$ increases the distance by one, because it adds different phases for different powers $m$ and $m'$.
 Therefore, for two different codewords, we have
 \begin{equation}
   \bra{\phi_m} E^\dagger F \ket{\phi_{m'}} =0 \qquad \text{if} \ \ \wt{(E^\dagger F )} < \ceil{n/2} \ .
 \end{equation}
 This is one of the Knill-Laflamme conditions Eq.~\eqref{eq:Knill-Laflamme}, in which two different codewords should be distinguishable in the presence of errors that act non-trivially on $\wt{(E^\dagger F)} < d$ sites.
 
 Moreover, errors $E$ and $F$ should not be able to change an encoded state for the weight $\wt{(E^\dagger F)} < d$, i.e., for two given codewords it is necessary to have $\bra{\phi_m} E^\dagger F \ket{\phi_m}=\bra{\phi_{m'}} E^\dagger F \ket{\phi_{m'}}$. 
 As the unitary operator $\modM$ is local, its application does not change the entanglement properties of a state, therefore all the  states $\ket{\phi_m}$ are AME states, then
 \begin{equation}
   \bra{\phi_m} E^\dagger F \ket{\phi_m} = \Tr{(E^\dagger F)} = 0 \qquad \forall m\in [\dloc] \ ,
 \end{equation}
 for $\wt{(E^\dagger F)} < \floor{n/2}$.
  In general, based on the Knill-Laflamme condition the subspace $\Cq = \lspan (\{{\ket{\phi_m} _{m \in [\dloc]}} \})$ is a QECC $[\![n,1,\floor{n/2}]\!]_\dloc$.
  \end{IEEEproof}
In Table \ref{table:mod-Sh-for-AME} we compare QECCs one can construct from AME state $\ket{\phi_0}$ Eq.\eqref{eq:AME-phi0},  using the Shortening and modified-Shortening processes.
We can see that the modified-Shortening provides quantum codes with smaller local dimension $\dloc$ given $n$ than previous codes.
With this method we also provide explicit codewords besides stabilizer formalism.

\begin{table*}[ht]
\begin{center}
\begin{tabular}[b]{ cccc} 
 & Shortening & \\
 \hline
 $\AME(n,\dloc)$ & $\longrightarrow$ & $[\![n-1,1, \floor{n/2}]\!]_\dloc$  & $\dloc$-dimensional Subspace \\
  \hline
   $[\![4 , 0, 3]\!]_{\dloc \geq 3}$ & $\longrightarrow$ & $[\![3 , 1, 2]\!]_{\dloc \geq 3}$ & $\AME(3,\dloc )$  \\ 
   $[\![5 , 0, 3]\!]_{\dloc \geq 4}$ & $\longrightarrow$ & $[\![4 , 1, 2]\!]_{\dloc \geq 4}$ & $1 \uni (4,\dloc)$ \\ 
   $[\![6 , 0, 4]\!]_{\dloc \geq 4}$ & $\longrightarrow$ & $[\![5 , 1, 3]\!]_{\dloc \geq 4}$ & $\AME(5,\dloc)$  \\ 
   $[\![7 , 0, 4]\!]_{\dloc \geq 7}$ & $\longrightarrow$ & $[\![6 , 1, 3]\!]_{\dloc \geq 7}$ & $2 \uni (6,\dloc)$ \\ 
   $[\![8 , 0, 5]\!]_{\dloc \geq 7}$ & $\longrightarrow$ & $[\![7 , 1, 4]\!]_{\dloc \geq 7}$ & $\AME(7,\dloc)$ \\ 
   $[\![9 , 0, 5]\!]_{\dloc \geq 8}$ & $\longrightarrow$ & $[\![8 , 1, 4]\!]_{\dloc \geq 8}$ & $3 \uni (8,\dloc)$ \\ 
   $\vdots$  & $\vdots$ & $\vdots$ & $\vdots$  \\
   $[\![n , 0, \floor{\frac{n}{2}}+1]\!]_{\dloc \geq n-1}$ & $\longrightarrow$ & $[\![n-1 , 1, \floor{\frac{n}{2}}]\!]_{\dloc \geq n-1}$ & $\floor{\frac{n-2}{2}} \uni (n-1,\dloc)$ \\ 
 \hline
\multicolumn{3}{c}{\vspace{0.5cm}}\\
 & modified-Shortening & \\
 \hline
 $\AME(n,\dloc)$ & $\longrightarrow$ & $[\![n,1, \floor{n/2}]\!]_\dloc$ & $\dloc$-dimensional Subspace  \\
  \hline
   $[\![4 , 0, 3]\!]_{\dloc \geq 3}$ & $\longrightarrow$ & $[\![4 , 1, 2]\!]_{\dloc \geq 3}$  & $\AME(4,\dloc)$ \\ 
   $[\![5 , 0, 3]\!]_{\dloc \geq 4}$ & $\longrightarrow$ & $[\![5 , 1, 2]\!]_{\dloc \geq 4}$ &$\AME(5,\dloc)$ \\ 
   $[\![6 , 0, 4]\!]_{\dloc \geq 4}$ & $\longrightarrow$ & $[\![6 , 1, 3]\!]_{\dloc \geq 4}$ & $\AME(6,\dloc)$ \\ 
   $[\![7 , 0, 4]\!]_{\dloc \geq 7}$ & $\longrightarrow$ & $[\![7 , 1, 3]\!]_{\dloc \geq 7}$ & $\AME(7,\dloc)$ \\ 
   $[\![8 , 0, 5]\!]_{\dloc \geq 7}$ & $\longrightarrow$ & $[\![8 , 1, 4]\!]_{\dloc \geq 7}$ & $\AME(8,\dloc)$ \\ 
   $[\![9 , 0, 5]\!]_{\dloc \geq 8}$ & $\longrightarrow$ & $[\![9 , 1, 4]\!]_{\dloc \geq 8}$ & $\AME(9,\dloc)$ \\ 
   $\vdots$  & $\vdots$ & $\vdots$ & $\vdots$  \\
   $[\![n , 0, \floor{\frac{n}{2}}+1]\!]_{\dloc \geq n-1}$ & $\longrightarrow$ & $[\![n , 1, \floor{\frac{n}{2}}]\!]_{\dloc \geq n-1}$ & $\AME(n,\dloc)$ \\ 
 \hline
 \end{tabular}
\end{center}
 \caption{\label{table:mod-Sh-for-AME} Comparison between code parameters and subspaces one can construct starting from $\AME(n,\dloc)$ state over $GF(\dloc)$, Eq.~\eqref{eq:AME-phi0}, using Shortening and modified-Shortening processes.}
\end{table*}

\section{Conclusions}
We have studied the remarkable relation between classical optimal codes, maximally multipartite entangled states, and quantum error correcting codes.
This study, in general, can lead to the construction of optimal quantum error correcting codes from highly entangled
subspaces.
We discussed a method that starts from a $k \uni$ state and by removing one party constructs a set of stabilizer QECCs.
Our construction provided the list of codewords besides presenting the stabilizer formalism.
Along the way, we have also shown that this method can be iterated and how to find the codewords in each step.
Then, we extended the connection between classical codes, $k \uni$ states and quantum codes to provide codes with larger code subspace compared with the existed constructions.
We have shown how to modify the method to produce
QECCs starting from an AME state without removing any party.
Our method, called the modified-Shortening construction, is explicit, physically motivated and works with a smaller
local dimension than previous codes. 
This has led to construct stabilizer QECCs $[\![ n,1,\floor{n/2}]\!]_\dloc$ starting from AME state or quantum code $[\![ n,0,\floor{n/2}+1]\!]_\dloc$.

It is an open question to check if the modified-Shortening could be applied recursively to achieve a larger set of QECCs starting from AME states.
A precise understanding of this method can encourage to  extend the modified-Shortening as a construction to produce $[\![n,k,\floor{n/2}+1-k]\!]_\dloc$ quantum codes.
This method and finding the set of codewords can also lead to having a better understanding of completely entangled subspaces (CESs) or genuinely entangled subspaces (GESs).

\section*{Acknowledgment}
The author would thank Maciej Demianowicz, Felix Huber, Sam Short-Morley,  Adam~Teixid\'{o}
as well as Antonio~Ac\'{i}n and Christian~Gogolin for very useful comments and insightful discussions.
I acknowledge support from the Spanish MINECO (Severo Ochoa SEV-2015-0522), Fundacio Cellex and Mir-Puig, Generalitat de Catalunya (SGR 1381 and CERCA Programme), and ERC AdG CERQUTE.

\appendices
\section{stabilizer formalism of the \lowercase{$k\uni$} states constructed from MDS codes.}\label{app:stab-kuni}
Let us first start with more details of the classical codes.
For a given linear classical code with parameters $[n,k,\dH]_\dloc$, a generator matrix is a $k\times n$ matrix, $G_{k\times n}$, over a finite field $GF(\dloc)$.
Given the generator matrix in standard form Eq.~\eqref{eq:generator-standard}, is useful to find a matrix which is called \emph{parity check} matrix $H_{(n-k)\times n}=[-A^T | \1_{n-k}]$.
The two matrices are related with $GH^T = 0$. 

Let $\C$ be an MDS code $[n,k,n-k+1]_\dloc$, over $GF(\dloc)$, the following statements for the generator matrix $G_{k\times n}$ and parity check matrix $H_{(n-k)\times n}$ are equivalent (see \cite[Chapter~11, Corollary 3 and Theorem 8]{MacWilliams} and \cite{ZCAA}).
\begin{Corollary} \label{Corollary:MDS-code}
 A linear code $\C$ is an MDS code if and only if
  \begin{itemize}
   \item[(i)]every square submatrix of $A$ is nonsingular.
   \item[(ii)]any subset of up to $k$ of the column vectors of $G_{k\times n} = [\1 |A]$ is linearly independent.
   \item[(ii)]any subset of up to $n-k$ of the column vectors of $H_{(n-k)\times n} = [-A^T |\1]$ is linearly independent
 \footnote{Parity check matrix $H_{(n-k)\times n}$ of a code $\C$ is generator matrix for its dual $\C^\perp$. In the case that the original code is an MDS code with parameters $\C = [n,k,n-k+1]_\dloc$, the dual code is also an MDS code with code parameters $\C^\perp = [n,n-k,\dH^{\perp}=k+1]_\dloc$. }.
 \end{itemize}
\end{Corollary}

The $k \uni$ state of minimal support constructed from MDS code, Eq.~\eqref{eq:kuni-MDScode}, recall
\begin{equation}
\ket{\psi} = \sum_{\vec{v}\in {GF(\dloc)}^k} \, \ket{\vec{v}\, G_{k\times n}}\ ,
\end{equation}
 is the plus one eigenstate of  $n$ stabilizer operators.
The generators are divided into two sets, $X$ stabilizers, $S_X$, and $Z$ stabilizers, $S_Z$,
\begin{equation}\label{eq:psi-stabilizers-MDS-G-H}
  S \coloneqq
  \begin{cases}
    S_X = \bigotimes_{j=1}^{n} X^{g_{i,j}} & 1 \leq i \leq k \\
    S_Z = \bigotimes_{j=1}^{n} Z^{h_{i,j}} & 1 \leq i \leq n-k
  \end{cases} \ ,
\end{equation}
where the matrix elements of $G_{k\times n}$ are denoted by $g_{i,j}$ and that of the code's parity check matrix $H_{(n-k) \times n}$ by $h_{i,j}$.
The first $k$ generators involve the $X$ operators (the $X$ stabilizers).
This forms a set of stabilizers, because adding the same codeword to all other codewords is just a relabeling of the terms in the summation.
Another set of stabilizers, $n-k$ of them, can be constructed from the $Z$ operators (the $Z$ stabilizers).
The action of product of stabilizers $S_Z$ leave state $\ket \psi$ invariant because of the fact that $G_{k\times n} (H_{(n-k)\times n})^T=0$, (see also \cite{ZCAA, Gottesman-thesis}).

\section{Stabilizers group of the code state space }\label{app:stab-Shortening}
The stabilizer formalism of the state $\ket{\psi _0}$, Eq.~\eqref{eq:psi0}, can be found by taking advantage of the connection to the classical coding theory.
Therefore, based on Eq.~\eqref{eq:psi-stabilizers-MDS-G-H}, one can find $n-1$ generators of the stabilizers of the state $\ket {\psi_0}$,
  \begin{equation}\label{eq:psi-stabilizers}
    S^{\psi_0} \coloneqq
     \begin{cases}
      \bigotimes_{j=1}^{n-1}\ X^{\Sh{g}_{i,j}} & 1 \leq i \leq k-1 \\
      \bigotimes_{j=1}^{n-1}\ Z^{\Sh{h}_{l,j}} & k \leq i \leq n-1
    \end{cases} \ ,
\end{equation}
where the matrix elements of $G_{\remove{k}}$ are denoted by $\Sh{g}_{i,j}$ and that of the code's parity check matrix by $\Sh{h}_{i,j}$.

For the code  $\Cq \coloneqq \lspan(\{\ket{\psi_m}_{m\in [\dloc]}\}) \subset \mathbb{C}_{\dloc}^{\otimes {n-1}}$
  with $\ket{\psi_m}=M^m \ket{\psi_0}$, Eq.~\eqref{eq:psi_m-Shortening} to be a stabilizer code, we need to generate a stabilizer group that stabilizes the given subspace.
  The set of the stabilizers $S^{\Cq}$ should satisfy the following equality
  \begin{equation}\label{eq:condition-stabilizers-shritening}
    \forall i,m \colon 
    \quad S^{\Cq}_i\,M^m\,\ket{\psi_0} = M^m\,\ket{\psi_0}\ .
  \end{equation}
  The above condition implies that every $S^{\Cq}_i\in S^{\Cq}$ must commute with $M$ (and hence $M^m$) operator and stabilize the state $\ket{\psi_0}$.  
The $M$ operator is a vector of exponents of the $X$ operators.
Therefore, the $k-1$ generators of the stabilizer group of the state $\ket{\psi_0}$ that involve $X$ operators, $S^{\psi_0}_X$, (first equation of Eq.~\eqref{eq:psi-stabilizers}), commute with $M$ and hence leave the state $\ket{\psi_m}$ invariant.
In order to find the stabilizers $S^{\Cq}_i$ that involve the $Z$ operators, we first consider direct computation for any two Pauli strings.
For two Pauli strings $A$ and $B$ the commutator follows 
  \begin{equation}
    A\,B = \omega^{\vec{A} \odot \vec{B}}\,B\,A \ ,
  \end{equation}
  where $\vec{A} = (\vec{A}_X,\vec{A}_Z)$ and $\vec{b}$ is defined in the same way and, 
   \begin{equation}
    \vec{A} \odot \vec{B} \coloneqq \vec{A}_Z \cdot \vec{B}_X - \vec{A}_X \cdot \vec{B}_Z \ .
  \end{equation} 
This implies that the stabilizers of $\ket{\psi_0}$ that involves $Z$ operators,$S_{Z}^{\psi_0}$ (the second equation of Eq.~\eqref{eq:psi-stabilizers}), satisfy the Eq.~\eqref{eq:condition-stabilizers-shritening} if for all $m$ it holds that $m \vec{m}_X \, \cdot \, \vec{S}^{\ \psi_0}_Z=0 \, \mod{\dloc}$.
This is also equivalent to just having $\vec{m}_X \cdot \vec{S}^{\ \psi_0}_Z=0 \, \mod{\dloc}$, where the vector $\vec{m}_X$ represent the vector of exponents 
in the $M$ operator.
The vector of exponents $S_{Z}^{\psi_0}$ of the $Z$ stabilizers is constructed from linear combination of the rows of the parity check matrix $\Sh H_{(n-k)\times(n-1)}$.  
Therefore, $\vec{S}^{\ \psi_0}_Z = \vec{v} \Sh H $, represent the vector of exponents of the $S_{Z}^{\psi_0}$, where $\vec{v} \in GF(\dloc)^{n-k}$.
The string of $Z$ operators that leave $\ket{\psi_m}=M^m\ket{\psi_0}$ invariant are those vector of exponents such that $\vec{m}_X \, . \, \vec{v} \Sh H =0$.
In general, the generator $S^{\Cq}$ of the stabilizer groups of $\Cq$ are 
 \begin{equation}\label{eq:psi-stabilizers}
    S^{\Cq} \coloneqq
     \begin{cases}
      \bigotimes_{j=1}^{n-1}\ X^{\Sh{g}_{i,j}} & 1 \leq i \leq k-1 \\
      \bigotimes_{j=1}^{n-1}\ Z^{\sum_{l=1}^{n-k} v_l\, \Sh{h}_{l,j}} & \text{where} \ \
      {\begin{array}{c}
   \vec{m}_X \, . \, \vec{v} \Sh H =0 \\
   \vec{v} \in GF(\dloc)^{n-k} \\
  \end{array} }
    \end{cases} \ .
    \end{equation}
The number of generators for the stabilizers group that involve $X$ operators is $k-1$ and those involve $Z$ operators is $n-k-1$, in total, they are $n-2$ generators.

\EOD


\begin{thebibliography}{00}
\bibitem{Steane96-multipartite-QECC}
A.~Steane, "Multiple Particle Interference and Quantum Error Correction",
Proc. Roy. Soc. Lond. A 452, 2551 (1996).

\bibitem{Rains}
E.~M. Rains, “Nonbinary quantum codes”,
 IEEE Transactions on Information Theory, 45, 1827 (1999).

\bibitem{Steane96}
A.~Steane, "Simple Quantum Error Correcting Codes", Phys. Rev. A 54, 4741 (1996).

\bibitem{Gottesman}
D.~Gottesman, 
"An Introduction to Quantum Error Correction and Fault-Tolerant Quantum Computation",
arXiv:0904.2557 (2009).

\bibitem{Calderbank1998}
A.~Calderbank, E.~Rains, P.~Shor, and N.~Sloane, “Quantum error correction via codes over GF(4)”, IEEE Trans. Inform. Theory, 44, 1369 (1998).

\bibitem{Scott2004}
A.~J. Scott, 
"Multipartite entanglement, quantum-error-correcting codes, and entangling power of quantum evolutions", 
Phys. Rev. A, 69, 052330, (2004).

\bibitem{ZCAA} 
Z. Raissi, C. Gogolin, A. Riera, A. Ac\'{i}n, 
"Constructing optimal quantum error correcting codes from absolute maximally entangled states",  
J. Phys A: Math. and Theor. 51, 075301 (2017), arXiv:1701.03359.

\bibitem{Gottesman-thesis}
D.~Gottesman,
"Stabilizer codes and quantum error correction",
arXiv:quant-ph/9705052 (1997).

\bibitem{ZACA}
Z.~Raissi, A.~Teixid\'{o}, C.~Gogolin, A.~Ac\'{i}n, "Constructing new k-uniform and absolutely maximally entangled states", arXiv:1910.12789.

\bibitem{Helwig-graphstates}
W.~Helwig,
"Absolutely Maximally Entangled Qudit Graph States",
arXiv:1306.2879 [quant-ph]

\bibitem{FelixMarkus}
F.~Huber, M.~Grassl, "Quantum Codes of Maximal Distance and Highly Entangled Subspaces" arXiv:1907.07733 [quant-ph]

\bibitem{Daniel}
D.~Alsina, M.~Razavi, "Absolutely maximally entangled states, quantum maximum distance separable codes, and quantum repeaters", arXiv:1907.11253 [quant-ph]

\bibitem{Sarvepalli}
P.~K. Sarvepalli, A.~Klappenecker, "Nonbinary Quantum Reed-Muller Codes", IEEE Proceedings International Symposium on Information Theory, ISIT (2005).

\bibitem{Ketkar}
A.~Ketkar, A.~Klappenecker, S.~Kumar, and P.~K. Sarvepalli, “Nonbinary stabilizer codes over finite fields", IEEE Trans. Info. Theory, 52, 4892, (2006).

\bibitem{Markus-Roetteler}
M.~Grassl, M.~Roetteler, "Quantum MDS Codes over Small Fields", IEEE International Symposium on Information Theory (ISIT) (2015) arXiv:1502.05267.

\bibitem{MacWilliams}
F.~J. MacWilliams, N.~J. A. Sloane, 
"The Theory of Error Correcting Codes",
North-Holland, Amesterdam (1977).

\bibitem{Singleton}
R. Singleton,
"Maximum distance q-nary codes", 
IEEE Trans. Inf. Theor., 10(2), 116, (2006).

\bibitem{Roth}
R. M. Roth, G. Seroussi, 
"On generator matrix of MDS codes",
IEEE Trans. Inform. Theory, vol. IT-31, pp. 826–830, (1985).

\bibitem{Seroussi}
G. Seroussi, R. M. Roth,
"On MDS extensions of generalized Reed-Solomon codes",
IEEE Trans. Inform. Theory, vol. IT-32, pp. 349–354, (1986).

\bibitem{Knill-Laflamme}
E.~Knill, R.~Laflamme, "A Theory of Quantum Error-Correcting Codes",
Physical Review A, 55, 900 (1997).

\bibitem{Grassl-on-optimal}
M.~Grassl, T.~Beth, M.~Roetteler, "On optimal quantum codes", International Journal of Quantum Information, 2, 55 (2004).

\bibitem{completely-entangled-subspaces}
K.~Parthasarathy, 
"On the maximal dimension of a completely entangled subspace for finite level quantum systems"
Proceedings Mathematical Sciences 114, 365 (2004).

\newpage

\end{thebibliography}
\end{document}